\documentstyle[pra,aps,12pt]{revtex}

\begin{document}
\draft
\author{Li-Bin Fu$^{1,}$ \thanks{%
Email: fu$_-$libin@mail.iapcm.ac.cn }, Jing-Ling Chen$^{1,2}$, and Xian-Geng
Zhao$^1$}
\address{$^1$Institute of Applied Physics and Computational Mathematics,\\
P.O. Box 8009 (28), 100088 Beijing, China\\
$^2$ Department of Physics, Faculty of Science, National University of\\
Singapore}
\title{Maximal violation of Clauser-Horne-Shimony-Holt inequality for two qutrits}
\maketitle

\begin{abstract}
Bell-Clauser-Horne-Shimony-Holt inequality (in terms of correlation
functions) of two qutrits is studied in detail by employing tritter
measurements. A uniform formula for the maximum value of this inequality for
tritter measurements is obtained. Based on this formula, we show that
non-maximally entangled states violate the Bell-CHSH inequality more
strongly than the maximally entangled one. This result is consistent with
what was obtained by Ac{\'{i}}n {\it et al} [Phys. Rev. A {\bf 65}, 052325
(2002)] using the Bell-Clauser-Horne inequality (in terms of probabilities).
\end{abstract}


\section{Introduction}

Bell inequality has come to be not only as a tool for exposing the weirdness
of quantum mechanics, but also as a more powerful resource in a number of
applications, such as in quantum communication.
Bell-Clauser-Horne-Shimony-Holt (Bell-CHSH) inequality has been applied in
communicating protocol (Ekert protocol) to detect the presence of the
eavesdropper \cite{ekert}. Furthermore, it has been found that two entangled
$N$-dimensional systems (qu$N$its) generate correlations that are more
robust against noise than those generated by two entangled qubits \cite
{chb1,chb2,chb3,gisin}. It was suggested that the higher dimensional
entangled systems may be much superior than two-dimensional systems in
quantum communication. Naturally, the extension of the protocol (involving
Bell-CHSH inequality) to higher dimension becomes an interesting problem.
So, it is necessary and important to investigate the Bell inequality for
higher dimensional systems.

In an interesting paper \cite{acin}, by using the Bell-Clauser-Horne
inequality (in terms of probabilities) \cite{chb3,gisin}, Ac{\'{i}}n {\it et
al}. have shown that non-maximally entangled states violate the Bell-CHSH
inequality more strongly than the maximally entangled one. Recently, a
Bell-CHSH inequality (in terms of correlation functions) of two qutrits has
been obtained \cite{cjl} by searching the inequality which can give the
minimal noise admixture $F_{thr}$ for the maximally entangled states. The
minimal noise admixture $F_{thr}$ for the maximally entangled state of two
qutrits has been obtained numerically by the method of linear optimization
in \cite{chb1} and analytically in \cite{gisin,fthr2}. The extension of the
Bell-CHSH to higher dimension is a non-trivial and interesting problem.
Actually, it has been applied to quantum cryptography \cite{dag}. In this
paper, we study the Bell-CHSH inequality of two qutrits for tritter
measurements by considering a class of pure states of two qutrits. A uniform
formula of the maximum value of this inequality is obtained. Based on this
formula, we find the states which give the maximum violation of the
Bell-CHSH inequality. This result is consistent with what was obtained by Ac{%
\'{i}}n {\it et al}. \cite{acin}.

\section{The inequality}

Let us consider a gedanken experiment with two observers each measuring two
observables on some state of two qutrits $\rho .$ We denote the observables
by $\widehat{A}^{i}$ $(i=1,2)$ for the first observer (Alice), $\widehat{B}%
^{j}(j=1,2)$ for the second observer (Bob). \ The measurement of each
observable yields three distinct outcomes which denote by $%
a_{1}^{i},a_{2}^{i}$ and $a_{3}^{i}$ for Alice's measurement of the
observable, and $b_{1}^{j},b_{2}^{j}$ and $b_{3}^{j}$ for Bob's measurement
of the observable. Specifically, the observables have the spectral
decompositions: $\widehat{A}^{i}=a_{1}^{i}\widehat{P}_{1}^{i}+a_{2}^{i}%
\widehat{P}_{2}^{i}+a_{3}^{i}\widehat{P}_{3}^{i},$ and $\widehat{B}%
^{j}=b_{1}^{j}\widehat{Q}_{1}^{j}+b_{2}^{j}\widehat{Q}_{2}^{j}+b_{3}^{j}%
\widehat{Q}_{3}^{j},$ where $\widehat{P}_{l}^{i}$ and $\widehat{Q}_{m}^{j}$ $%
(l,m=1,2,3)$ are mutually orthogonal projectors respectively. The
probability of obtaining the set of three numbers $(a_{l}^{i},b_{m}^{j})$ in
a simultaneous measurement of observables $\widehat{A}^{i}$ and $\widehat{B}%
^{j}$ on the state $\rho $ is denoted by $P(a_{l}^{i},b_{m}^{j}),$ which can
be given by the standard formula
\begin{equation}
P(a_{l}^{i},b_{m}^{j})=Tr(\rho \widehat{P}_{l}^{i}\otimes \widehat{Q}%
_{m}^{j}).  \label{pp}
\end{equation}

As introduced and used in Ref. \cite{first}, the correlation function $Q(%
\vec{\varphi}^{A_{i}},\varphi ^{B_{j}})$ ($Q_{ij}$ for short) between
Alice's and Bob's measurements is
\begin{equation}
Q_{ij}=\sum\limits_{l_{i},m_{j}=1}^{3}\alpha
^{l_{i}+m_{j}}P(a_{l_{i}},b_{m_{j}}),  \label{a}
\end{equation}
where $\alpha =e^{i2\pi /3}.$ Let us define the following quantity
\begin{equation}
S={\rm Re}[Q_{11}+Q_{12}-Q_{21}+Q_{22}]+\frac{1}{\sqrt{3}}{\rm Im}%
[Q_{11}-Q_{12}-Q_{21}+Q_{22}].  \label{s}
\end{equation}
It can be shown \cite{cjl}, using the recently discovered Bell inequality
for two qutrits \cite{chb3}, that according to local realistic theory $S$
can not exceed $2,$ i.e. $S\leq 2$ for local realistic theory. However, when
using the quantum correlation function given in Eq. (\ref{a}), $S_{\max }$
acquires the value $\frac{2}{9}(6+4\sqrt{3})\approx 2.87293$ for the state $%
\left| \psi \right\rangle =\frac{1}{\sqrt{3}}\sum\limits_{i}^{3}\left|
i\right\rangle \left| i\right\rangle $ $,$ the maximally entangled state$.$
Following \cite{chb1}, we define the threshold noise admixture $F_{thr}$
(the minimal noise admixture fraction for $\left| \psi \right\rangle $) $%
F_{thr}=1-2/S_{\max }.$ Then for the maximally entangled two qutrits, we
have $F_{thr}=0.30385.$ For the maximally entangled two qubits, one has $%
F_{thr}=0.29289.$ Obviously, entangled qutrits are more resistant to noise
than entangled qubits \cite{chb1,fthr2}.

As suggested in Ref. \cite{cjl} and \cite{chb3}, the Bell-CHSH inequality
for two qutrits can be expressed as
\begin{equation}
-4\leq S\leq 2.  \label{ine}
\end{equation}

On the other hand, the interesting thing is the maximal $F_{thr}$ of two
qutrits obtained in Ref. \cite{ghz} by the numerical linear optimization
method. The authors found that the optimal non-maximally entangled state of
two qutrits is around $3\%$ more resistant to noise than the maximally
entangled one. The maximal $F_{thr}=0.31386$ for such state (a non-maximally
state). Similar result is obtained in Ref.\cite{acin}. Obviously, the
maximal violation of the inequality should be $2.91485$ for such
non-maximally entangled states.

For simplicity, we consider such a gedanken experiment that Alice's and
Bob's observables are defined by unbiased symmetric six-port beam-splitter
on the state of two qutrits
\begin{equation}
\left| \psi \right\rangle =\frac{1}{\sqrt{3}}\sum\limits_{i}^{3}a_{i}\left|
i\right\rangle \left| i\right\rangle ,  \label{psi}
\end{equation}
with real coefficients $a_{i},$the kets $\left| i\right\rangle $ $(i=1,2,3)$
denote the orthonormal basis states for the qutrit. The unbiased symmetric
six-port beam-splitter, called tritter \cite{trit,chb7}, is an optical
device with three input and output ports. In front of every input port there
is a phase shifter that changes the phase of the photon entering the given
port. The observers select the specific local observables by setting
appropriate phase shifts in the beams leading to the entry ports of the
beam-splitters. Such process performs a unitary transformation between
``mutually unbiased'' bases in the Hilbert space \cite{bell8,bell9,bell10}.
The overall unitary transformation performed by such a device is given by
\begin{equation}
U_{ij}=\frac{1}{\sqrt{3}}\alpha ^{(i-1)(j-1)}e^{i\varphi _{j}},\;\;i,j=1,2,3
\label{u}
\end{equation}
where $\alpha =e^{i2\pi /3}$ and $j$ denotes an input beam to the device,
and $i$ an output one; $\varphi _{j}$ are the three phases that can be set
by the local observer, denoted as $\vec{\varphi}=(\varphi _{1},\varphi
_{2},\varphi _{3})$ . The transformations at Alice's side are denoted as $%
\vec{\varphi}^{A}=(\varphi _{1}^{A},\varphi _{2}^{A},\varphi _{3}^{A}),$ and
$\vec{\varphi}^{B}=(\varphi _{1}^{B},\varphi _{2}^{B},\varphi _{3}^{B})$ for
Bob's side.

The observables measured by Alice and Bob are now defined as follows. \ The
set of projectors for Alice's $i$-th measurement is given by $\widehat{P}%
_{l}^{i}=U_{A}^{+}(\vec{\varphi}^{A_{i}})\left| l\right\rangle \left\langle
l\right| U_{A}(\vec{\varphi}^{A_{i}})$ $(l=1,2,3),$ where $U_{A}(\vec{\varphi%
}^{A_{i}})$ is the matrix of Alice's unbiased symmetric six-port
beam-splitter defined by Eq. (\ref{u}). Bob's $j$-th measurement is given by
$\widehat{Q}_{m}^{j}=U_{B}^{+}(\vec{\varphi}^{B_{j}})\left| m\right\rangle
\left\langle m\right| U_{B}(\vec{\varphi}^{B_{j}})$ $(m=1,2,3).$ Then, from (%
\ref{pp}) and (\ref{a}), the correlation function for state $\left| \psi
\right\rangle $ reads
\begin{equation}
Q_{ij}=\sum\limits_{n,k}^{3}\sum\limits_{l_{i},m_{j}}^{3}a_{n}a_{k}\alpha
^{l_{i}+m_{j}}(\alpha ^{\ast })^{(n-1)(l_{i}+m_{j}-2)}\alpha
^{(k-1)(l_{i}+m_{j}-2)}e^{i(\varphi _{k}^{A_{i}}+\varphi
_{k}^{B_{j}}-\varphi _{n}^{A_{i}}-\varphi _{n}^{B_{j}})}.  \label{qq}
\end{equation}
This shows the results of the measurement obtained by Alice and Bob are
strictly correlated.

In the following, we will investigate the Bell-CHSH inequality (\ref{ine})
for the tritter measurements and give analytical discussions of above
results.

\section{The maximal violation}

By substituting Eq. (\ref{qq}) into (\ref{s}), after some elaborate, we
obtain
\begin{equation}
S=a_{1}a_{2}T_{12}+a_{1}a_{3}T_{13}+a_{2}a_{3}T_{23},  \label{ss}
\end{equation}
where
\[
T_{12}=\frac{1}{9}[3\cos (\varphi _{1}^{A_{2}}-\varphi _{2}^{A_{2}}+\varphi
_{1}^{B_{1}}-\varphi _{2}^{B_{1}})-3\cos (\varphi _{1}^{A_{1}}-\varphi
_{2}^{A_{1}}+\varphi _{1}^{B_{1}}-\varphi _{2}^{B_{1}})
\]
\[
-3\cos (\varphi _{1}^{A_{2}}-\varphi _{2}^{A_{2}}+\varphi
_{1}^{B_{2}}-\varphi _{2}^{B_{2}})-\sqrt{3}\sin (\varphi
_{1}^{A_{2}}-\varphi _{2}^{A_{2}}+\varphi _{1}^{B_{1}}-\varphi _{2}^{B_{1}})
\]
\[
+\sqrt{3}\sin (\varphi _{1}^{A_{1}}-\varphi _{2}^{A_{1}}+\varphi
_{1}^{B_{1}}-\varphi _{2}^{B_{1}})+2\sqrt{3}\sin (\varphi
_{1}^{A_{1}}-\varphi _{2}^{A_{1}}+\varphi _{1}^{B_{2}}-\varphi _{2}^{B_{2}})
\]
\begin{equation}
+\sqrt{3}\sin (\varphi _{1}^{A_{2}}-\varphi _{2}^{A_{2}}+\varphi
_{1}^{B_{2}}-\varphi _{2}^{B_{2}})],  \label{t12}
\end{equation}
\[
T_{13}=-\frac{1}{9}[3\cos (\varphi _{1}^{A_{1}}-\varphi _{3}^{A_{1}}+\varphi
_{1}^{B_{1}}-\varphi _{3}^{B_{1}})-3\cos (\varphi _{1}^{A_{2}}-\varphi
_{3}^{A_{2}}+\varphi _{1}^{B_{1}}-\varphi _{3}^{B_{1}})
\]
\[
+3\cos (\varphi _{1}^{A_{2}}-\varphi _{3}^{A_{2}}+\varphi
_{1}^{B_{2}}-\varphi _{3}^{B_{2}})+\sqrt{3}\sin (\varphi
_{1}^{A_{1}}-\varphi _{3}^{A_{1}}+\varphi _{1}^{B_{1}}-\varphi _{3}^{B_{1}})
\]
\[
-\sqrt{3}\sin (\varphi _{1}^{A_{2}}-\varphi _{3}^{A_{2}}+\varphi
_{1}^{B_{1}}-\varphi _{3}^{B_{1}})+2\sqrt{3}\sin (\varphi
_{1}^{A_{1}}-\varphi _{3}^{A_{1}}+\varphi _{1}^{B_{2}}-\varphi _{3}^{B_{2}})
\]
\begin{equation}
+\sqrt{3}\sin (\varphi _{1}^{A_{2}}-\varphi _{3}^{A_{2}}+\varphi
_{1}^{B_{2}}-\varphi _{3}^{B_{2}})],  \label{t13}
\end{equation}
and
\[
T_{23}=-\frac{1}{9}[3\cos (\varphi _{2}^{A_{1}}-\varphi _{3}^{A_{1}}+\varphi
_{2}^{B_{1}}-\varphi _{3}^{B_{1}})-3\cos (\varphi _{2}^{A_{2}}-\varphi
_{3}^{A_{2}}+\varphi _{2}^{B_{1}}-\varphi _{3}^{B_{1}})
\]
\[
+3\cos (\varphi _{2}^{A_{2}}-\varphi _{3}^{A_{2}}+\varphi
_{2}^{B_{2}}-\varphi _{3}^{B_{2}})-\sqrt{3}\sin (\varphi
_{2}^{A_{1}}-\varphi _{3}^{A_{1}}+\varphi _{2}^{B_{1}}-\varphi _{3}^{B_{1}})
\]
\[
+\sqrt{3}\sin (\varphi _{2}^{A_{2}}-\varphi _{3}^{A_{2}}+\varphi
_{2}^{B_{1}}-\varphi _{3}^{B_{1}})-2\sqrt{3}\sin (\varphi
_{2}^{A_{1}}-\varphi _{3}^{A_{1}}+\varphi _{2}^{B_{2}}-\varphi _{3}^{B_{2}})
\]
\begin{equation}
-\sqrt{3}\sin (\varphi _{2}^{A_{2}}-\varphi _{3}^{A_{2}}+\varphi
_{2}^{B_{2}}-\varphi _{3}^{B_{2}})],  \label{t23}
\end{equation}
are three continuous functions of twelve angles $\vec{\varphi}^{A_{i}}$ and $%
\vec{\varphi}^{B_{j}}$ $(i,j=1,2).$ So, $S$ is the continuous function of
the twelve variables. The points which satisfy
\begin{equation}
\frac{\partial S}{\partial \varphi _{j}^{\Lambda _{i}}}=0,\text{ }\Lambda
=A,B;\;i=1,2;\;and\;j=1,2,3,  \label{cond}
\end{equation}
are the critical points of the function $S$. According to the theory of
extreme points of continuous functions, we know that the extreme points are
belong to the critical points of the function. So, we can extract the
maximum and minimum of $S$ from the critical points by comparing the value
of $S$ among the critical points, since the maximum and minimum point must
be one of extreme points.

On the other hand, we can know that $|t_{12}|\leq \frac{4}{3},|t_{13}|\leq
\frac{4}{3},|t_{23}|\leq \frac{4}{3}.$ However, the above three formulae are
strongly correlated, so $t_{12},$ $t_{13},$ and $t_{23}$ can not reach their
maximum value at the same time. It happens that when one of $t_{12},$ $%
t_{13},$ and $t_{23}$ reaches its maximum value $\frac{4}{3},$ the others
can reach their sub-maximum value $\frac{4}{3\sqrt{3}}.$ If we consider $%
t_{12},$ $t_{13},$ and $t_{23}$ as three coordinates, then they can form a
complicated polyhedron. The polyhedral vertices are the points when $t_{12},$
$t_{13},$ and $t_{23}$ reach their extreme values.

{\bf Lemma} For the formula $G=\sum_{i=1}^N\xi _iR_i$ $,$ where $\xi _i$ are
$N$ real parameters, the maximum (minimum) points of $G$ must on the
boundary of the region formed by $R_i$ for any $\xi _i.$

Proof: Giving $G^{0}=\sum_{i=1}^{N}\xi _{i}$ $R_{i}^{0}$, if $%
R_{i}^{0}(i=1,2,\cdots ,N)$ are in the inner region formed by $R_{i}$, we
can always have $G=G^{0}+\sum_{i=1}^{N}\xi _{i}\Delta R_{i},$ in which $%
\Delta R_{i}$ are infinitesimal values satisfying $\xi _{i}\Delta R_{i}>0$ $%
(i=1,2,\cdots ,N),$ so that $G>G^{0};$ or $\Delta R_{i}$ are infinitesimal
values satisfying $\xi _{i}\Delta R_{i}<0$ $(i=1,2,\cdots ,N),$ so that $%
G<G^{0}.$ So, we can know that the maximum (minimum) points of $G$ can only
find on the boundary.

{\bf Theorem} The maximum and minimum value of $S$ for a given state (\ref
{psi}) must be found at the vertices of polyhedron formed by $t_{ij}$ $%
(i\neq j,i,j=1,2,3).$

Proof: We know that the maximum points of $S$ is belong to the critical
points of $S.$ For the critical points in the inner region formed by $%
(t_{12},t_{13},t_{23}),$ from the {\bf Lemma} we know that the value of such
critical points must be less than some values of $S$ on the boundary, so
they can not be the maximum points of $S$. For the same reason, if the
critical point on the boundary (excepting for vertices), we can know that
the value of $S$ on this point must be less than $S$ on one of the vertices
on this boundary. Then, the maximum value of $S$ must be only found on the
vertices of region formed by $(t_{12},t_{13},t_{23})$.

In analog to the above discussion, the minimum value of $S$ can also be
found on the vertices.

To find out the maximum (minimum) value we have to calculate the vertices of
the polyhedron formed by $t_{ij}.$ For convenience, we denote $T_{1}\ $as
one of $\{t_{12},t_{13},t_{23}\},$ $T_{2}$ as one of $\{t_{12},t_{13},t_{23}%
\}/\{T_{1}\}$ and $T_{3}$ as one of $\{t_{12},t_{13},t_{23}\}/\{T_{1},T_{2}%
\},$where $\{\}/\{\}$ means division of sets namely, if $T_{1}=t_{12}$, then
$T_{2}\in \{t_{12},t_{13},t_{23}\}/\{t_{12}\}=\{t_{13},t_{23}\},$ and so on.
In the following, we list the vertices formed by the maximum and sub-maximum
of $t_{ij}$ (it is enough),
\begin{equation}
(|T_{1}|,|T_{2}|,|T_{3}|)=(\frac{4}{3},\frac{4}{3\sqrt{3}},\frac{4}{3\sqrt{3}%
}),\text{ }for\text{ }\;T_{1}T_{2}T_{3}>0;  \label{tt123}
\end{equation}
and
\begin{equation}
(|T_{1}|,|T_{2}|,|T_{3}|)=(\frac{4}{3},\frac{4}{3},\frac{4}{3}),\text{ }for%
\text{ }\;T_{1}T_{2}T_{3}<0.  \label{ttt1234}
\end{equation}
Comparing the value of $S$ among these points, we can obtain the maximum and
minimum values of $S$ for the state (\ref{psi}). Assuming $\{K_{i},\
(i=1,2,3)\}=$ $\{|a_{1}a_{2}|,|a_{1}a_{3}|,|a_{2}a_{3}|\}$ $,$ where $``="$
means the equality of two sets, and $K_{i}$ are in decreasing order, i.e. $%
K_{1}\geq K_{2}\geq K_{3},$ let us define
\begin{equation}
S_{1}(\left| \psi \right\rangle )=\frac{4}{3}K_{1}+\frac{4}{3\sqrt{3}}%
(K_{2}+K_{3}),  \label{ss1}
\end{equation}
and
\begin{equation}
S_{2}(\left| \psi \right\rangle )=\frac{4}{3}(K_{1}+K_{2}-K_{3}),
\label{ss2}
\end{equation}
Then, we can know the maximum value of $S$ must be
\begin{equation}
S_{\max }(\left| \psi \right\rangle )=Max(S_{1}(\left| \psi \right\rangle
),S_{2}(\left| \psi \right\rangle )).  \label{ssmax}
\end{equation}
From (\ref{ss1}) and (\ref{ss2}), we know that $S_{2}(\left| \psi
\right\rangle )\geq S_{1}(\left| \psi \right\rangle )$ only for $\frac{K_{3}%
}{K_{2}}\leq 2-\sqrt{3}.$ If taking $\sum_{i}a_{i}^{2}=3$ into account, one
can prove that when $Max(|a_{1}|,|a_{2}|,|a_{3}|)\geq \frac{\sqrt{6+3\sqrt{3}%
}}{2}=1.67303$ , $S_{2}(\left| \psi \right\rangle )\geq S_{1}(\left| \psi
\right\rangle ).$ Let us define $A_{\max }=Max(|a_{1}|,|a_{2}|,|a_{3}|),$
finally we obtain that
\begin{equation}
S_{\max }(\left| \psi \right\rangle )=\left\{
\begin{array}{cc}
\frac{4}{3}K_{1}+\frac{4}{3\sqrt{3}}(K_{2}+K_{3}), & \;A_{\max }\leq \frac{%
\sqrt{6+3\sqrt{3}}}{2}; \\
\frac{4}{3}(K_{1}+K_{2}-K_{3}), & \;A_{\max }>\frac{\sqrt{6+3\sqrt{3}}}{2}.
\end{array}
\right.  \label{ssmaxx}
\end{equation}
We can also prove that the minimum of $S$ is
\begin{equation}
S_{\min }(\left| \psi \right\rangle )=-\frac{4}{3}(K_{1}+K_{2}+K_{3}).
\label{ssmin}
\end{equation}
Obviously one can easily find that for maximally entangled state $\left|
\psi \right\rangle =\frac{1}{\sqrt{3}}\sum\limits_{i}^{3}\left|
i\right\rangle \left| i\right\rangle $, (i.e., $a_{i}=1$), we have $S_{\max
}=\frac{2}{9}(6+4\sqrt{3})$ and $S_{\min }=-4,$ which are the same as the
results obtained in Refs. \cite{chb3,acin,ghz,cer}$.$

In Fig.1, we give the comparison between the theoretical results and the
numerical calculations obtained by multi random search optimization method,
which shows a perfect agreement; (a) for $S_{\max }$ and (b) for $S_{\min },$
in which $a_{1}$ changes in region $[-\sqrt{3},\sqrt{3}]$, $a_{2}=\sqrt{%
(3-a_{1}^{2})\varepsilon }$ and $a_{3}=\sqrt{(3-a_{1}^{2})(1-\varepsilon )},$
$0\leq \varepsilon \leq 1$. One can find some inflexion points in fig. 1(a),
for example, at the point $a_{1}=1$ when $\varepsilon =0.5.$ These inflexion
points are due to the discontinuous change of $K_{1},$ the maximum value
among $|a_{1}a_{2}|,|a_{1}a_{3}|$ and $|a_{2}a_{3}|,$ e.g., for $\varepsilon
=0.5,$ $K_{1}=a_{2}a_{3}=(\frac{3-a_{1}^{2}}{2})$ when $a_{1}\leq 1,$ but
when $a_{1}>1$, $K_{1}=a_{1}a_{2}=a_{1}\sqrt{\frac{(3-a_{1}^{2})}{2}}.$ On
the other hand, we can see from Fig.1(a) that the maximally entangled states
are not the states that give the maximal violation of the Bell inequality.

Consider $a_{i}$ as variables, we can obtain the maximal value of $S_{\max
}, $ denoted as $\bar{S}_{\max },$ by calculating the extreme value of Eq. (%
\ref{ssmaxx}), after some elaboration, we get
\begin{equation}
\bar{S}_{\max }=1+\sqrt{\frac{11}{3}},  \label{smax1}
\end{equation}
when
\begin{equation}
\{|a_{1}|,|a_{2}|,|a_{3}|\}=\{\sqrt{\frac{3}{2}\left( 1-\sqrt{\frac{3}{11}}%
\right) },\;\sqrt{\frac{3-a_{1}^{2}}{2}},\sqrt{\frac{3-a_{1}^{2}}{2}}\}.
\label{a1}
\end{equation}
One sees that for this value the threshold amount of noise is about $%
F_{thr}=0.3139,$ which is as the same as what has been obtained in recently
calculation \cite{acin,ghz,cer}. So, this result gives another evidence for
the inequality (\ref{ine}).

On the other hand, we can also calculated the minimum value of $S_{\min },$
denoted as $\bar{S}_{\min }$
\begin{equation}
\bar{S}_{\min }=-4,\text{ }for\text{ }\{|a_{1}|,|a_{2}|,|a_{3}|\}=\{1,\;1,1%
\}.  \label{mmmin}
\end{equation}
Then, we can know that
\begin{equation}
0\leq S_{\max }\leq 1+\sqrt{\frac{11}{3}},\;-4\leq S_{\min }\leq 0.
\label{region}
\end{equation}
Obviously, for tritter measurements, the left hand of the inequality (\ref
{ine}) would never be violated, and the right hand only be violated by some
of pure states. We can easily find the states that violate the inequality
for tritter measurements from the formula (\ref{ssmaxx}). In Fig. 2, we show
the states described by $(a_{1},a_{2}=\sqrt{(3-a_{1}^{2})\varepsilon }%
,\;a_{3}=\sqrt{(3-a_{1}^{2})(1-\varepsilon )}\;)$ that violate the
inequality for tritter measurements$.$ The states which violate the
inequality are in the shadow region; the states of which $S_{\max }=2$ are
on the boundary of the shadow region; the states in other region can not
violate the inequality for tritter measurements.

We should add here that some similar calculations as well as some
equivalence results were made by Cereceda \cite{cer} where the author
compared some of the two-qutrit inequalities and investigated them in detail.

\section{Discussion}

In the above discussion we only concentrate on tritter measurements which
can be easily carried out for nowadays technology \cite{trit}. By detail
studying the Bell-CHSH inequality of two qutrits, we give formulae of the
maximum and minimum values of this inequality, and obtain the states which
give the maximal violation of the Bell-CHSH inequality. The maximal
violation we obtained are the same as Refs. \cite{acin,ghz} .

Indeed, one should use general measurements to study the problem of
maximizing the Bell violation for a state, or in other words, for some
states the tritter measurements are not optimal.

So, some states that do not violate the inequality using tritter
measurements, but may violate the inequality when general measurements are
taken into account \cite{cer}. For example, for the state with $|a_{1}|=1.56$
and $\varepsilon =0.5$, $S_{\max }=1.964$ for tritter measurements, which
does not violate the inequality; but if we employ the following
measurements, $\widehat{P}_{l}^{i}=U_{A}^{+}(\vec{\varphi}^{A_{i}})\left|
x_{l}\right\rangle \left\langle x_{l}\right| U_{A}(\vec{\varphi}^{A_{i}})$ $%
(l=1,2,3)$ and $\widehat{Q}_{m}^{j}=U_{B}^{+}(\vec{\varphi}^{B_{j}})\left|
x_{m}\right\rangle \left\langle x_{m}\right| U_{B}(\vec{\varphi}^{B_{j}})$ $%
(m=1,2,3)\ $where $\left| x_{1}\right\rangle =\frac{1}{\sqrt{2}}[\left|
1\right\rangle +\left| 2\right\rangle ],$ $\left| x_{2}\right\rangle =\frac{1%
}{\sqrt{2}}[\left| 1\right\rangle -\left| 2\right\rangle ]$ and $\left|
x_{3}\right\rangle =\left| 3\right\rangle $ are orthonormal basis, we can
obtain $S_{\max }=2.0132$ (violates the inequality).

However, by employing tritter measurements, it can reveal many important
properties of Bell inequality of entangled two qutrits. For instance, for
the maximally entangled state $\left| \psi \right\rangle =\frac{1}{\sqrt{3}}%
\sum\limits_{i}^{3}\left| i\right\rangle \left| i\right\rangle $ and the
states that maximally violates the inequality, the tritter measurements are
optimal, and based on such entangled qutrit pairs a cryptographic protocol
has been presented more recently \cite{dag} by employing tritter
measurements.

\section{Acknowledgment}

We thank Professor S.G. Chen for useful discussions. This work was supported
by the 973 Project of China and Science and Technology Funds of CAEP.

\section{Figures caption:}

Fig. 1(a). The maximal value of the inequality for tritter measurements, $%
S_{\max },$ for state given by Eq. (\ref{psi}), where $a_{1}$ changes in
region $[-\sqrt{3},\sqrt{3}]$, $a_{2}=\sqrt{(3-a_{1}^{2})\varepsilon }$ and $%
a_{3}=\sqrt{(3-a_{1}^{2})(1-\varepsilon )},$ $0\leq \varepsilon \leq 1$. The
solid lines are theoretical results, circles are numerical dates; dotted
line shows the maximal value predicted by the local realistic theory, dashed
line marks the value of the maximally entangled states. (b) The minimal
value of the inequality, $S_{\min }$.

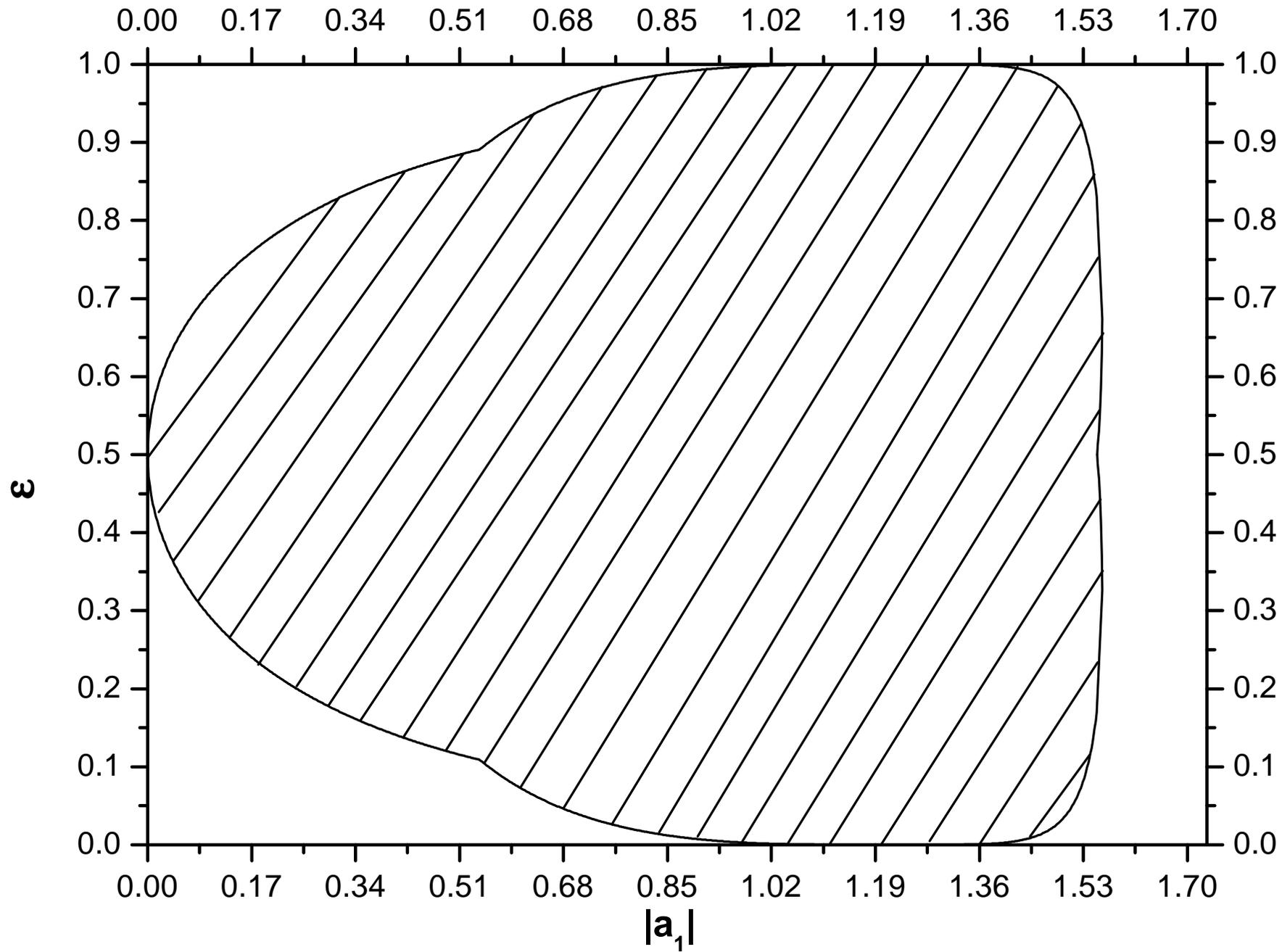
Fig. 2. It shows the states that violate the inequality for tritter
measurements. The states in shadow region violate the inequality.

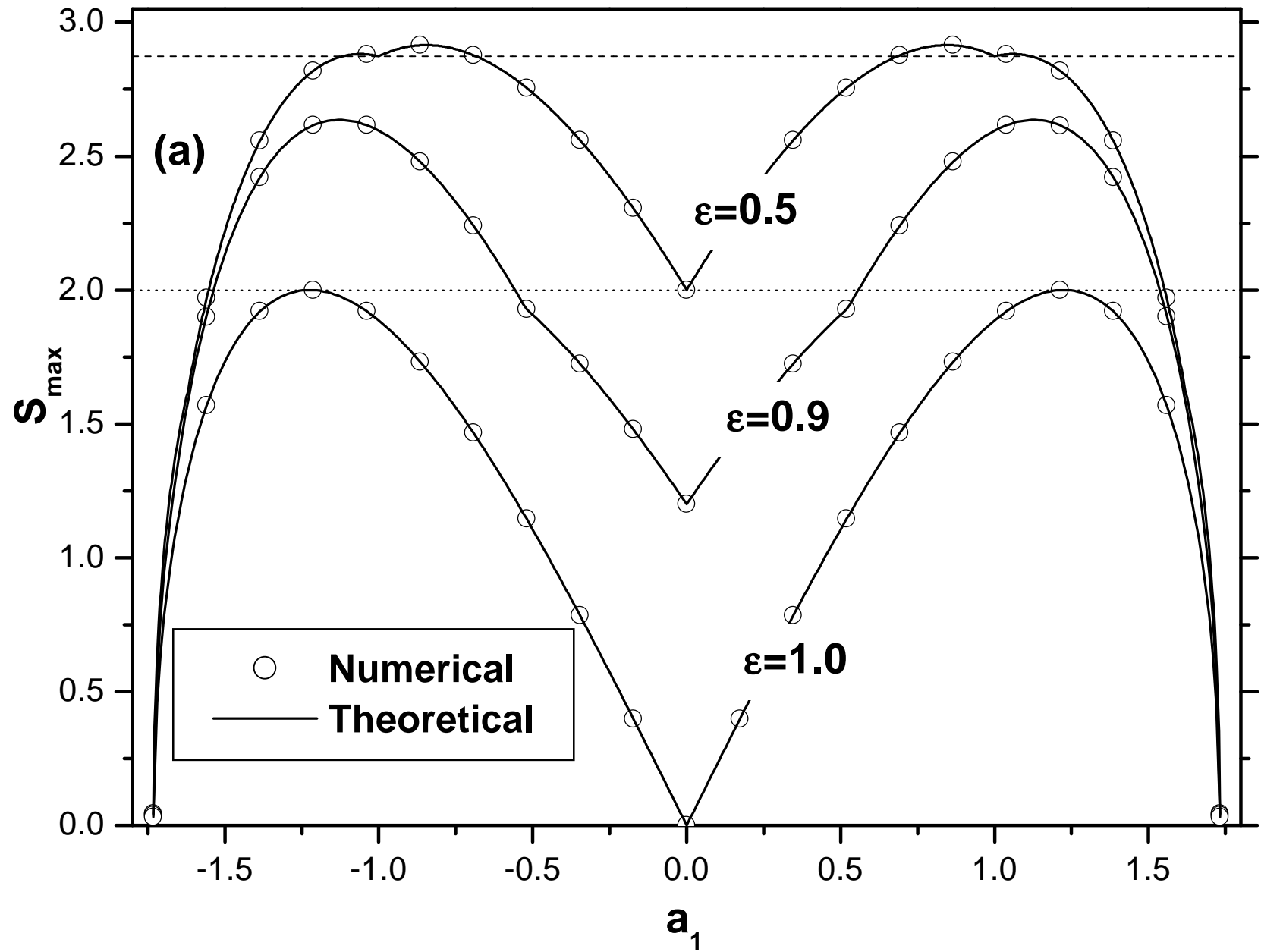
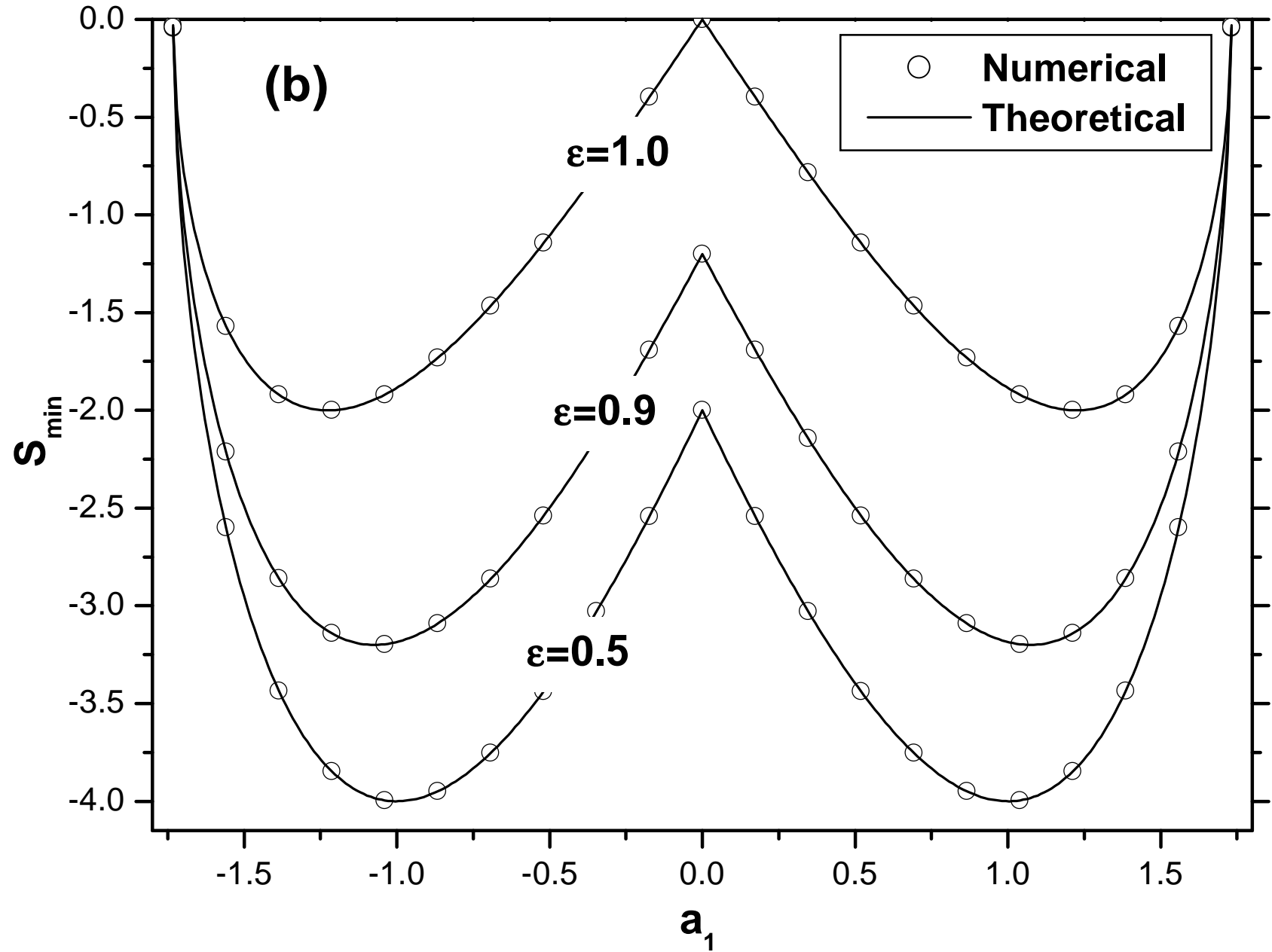
\end{document}